%Paper: gr-qc/9409056
%From: Dalia Goldwirth <dalia@fnas09.fnal.gov>
%Date: Tue, 27 Sep 94 10:04:41 -0600

\magnification\magstep1
\font\cs=cmr10 scaled \magstep3
\vskip 1.5 true cm
\centerline{\cs Conditions for inflation}
\vskip 1 true cm
\centerline{\cs in an initially inhomogeneous universe}
\vskip 1.5 true cm
\centerline{Nathalie Deruelle}
\smallskip
\centerline{\it D\'epartement d'Astrophysique Relativiste et de Cosmologie,}
\centerline{\it Centre National de la Recherche Scientifique, Observatoire de
Paris, 92195 Meudon, France}
\centerline{\it and Department of Applied Mathematics and Theoretical Physics,}
\centerline{\it University of Cambridge, Silver Street, Cambridge CB3 9EW,
England}
\bigskip
\centerline{Dalia S. Goldwirth}
\smallskip
\centerline{\it School of Physics and Astronomy, Raymond and Beverly Sackler}
\centerline{\it Faculty of Exact Sciences, Tel-Aviv University, 69976}
\centerline{\it Tel-Aviv, Israel}
\vskip 1cm
%\openup 3\jot
\bigskip
\centerline{8 September 1994}

\vskip 2cm
{\bf Abstract}
\medskip

 Using a long wavelength iteration scheme to solve Einstein's
equations near the Big-Bang singularity of a universe driven by a massive
scalar field,
we find how big initial quasi-isotropic inhomogeneities can be before they can
prevent inflation to set in.

\vfill\eject

\vskip 0.5 cm
\noindent
{\bf 1. Introduction}
\vskip 0.5 cm

The scale factor $a(t)$ of a spatially flat isotropic and homogeneous
Robertson-Walker (fRW) universe driven by a massive scalar field $\varphi$
evolves as follows: near the Big-Bang (chosen to be $t=0$) the scalar field
goes
to $\infty$ as $-\ln t$ and behaves like a perfect stiff fluid  (whose pressure
equals the energy density, that is whose  adiabatic index $\Gamma$ is 2), so
that $a$ grows as $t^{1/3}$ [1]. Then $a$ grows quasi exponentially during the
inflationary regime during which $\varphi$ slowly rolls down its potential well
[2]. Finally at the end of inflation when $\varphi$ oscillates in the bottom of
the well $a$ behaves as if the universe was driven by dust and grows on
average as $t^{2/3}$ [3].

A question is : how stable is this evolution against departures from spatial
flatness, isotropy, and homogeneity. The effect of curvature was studied in
[4];
since its role after the Big-Bang can become predominant, it can  prevent
inflation from ever starting if strong enough. Inflation in
homogeneous albeit anisotropic Bianchi models was thoroughly analyzed, with
similar conclusions : see e.g. [5] for a review.  Finally, the role of
inhomogeneities was studied analytically under simplifying assumptions in [6].
They were studied numerically, in the case of planar symmetry in [7] and in the
case of  spherical symmetry in [8].  In [8], setting the initial conditions in
such a way that inflation would  occur only in the central region if the
universe evolved like a fRW space time, the conclusion was that inflation would
indeed occur only if the central region was larger than a few times the local
Hubble radius.

Here we shall tackle the problem semi-analytically in the long wavelength
approximation.

The long wavelength iteration scheme, the history of which goes back to [9],
is a way to build, out of  ``seed"
spatial metrics, approximate solutions of Einstein's equations which
describe inhomogeneous but quasi-isotropic universes on scales larger than the
local Hubble
radius (see [10] and references therein for a detailed description of the
scheme). When matter satisfies the strong energy condition (that is does not
inflate) this approximation is well suited to describe the early universe
since most scales are then larger than the Hubble radius on average (this is
the well-known horizon problem). These approximate solutions however are not
generic in the sense that they are built out of a seed spatial metric, that is
on three physically distinct arbitrary functions, instead of 6 (4 for the
gravitational field, plus 2 for the scalar field). The three missing functions
can be seen as describing departures from isotropy. These anisotropies cannot
be neglected near the Big-Bang (see e.g.
[9-10]) but they decay with time much faster than all other contributions
to inhomogeneity and hence will be ignored (see [11] for an analysis of the
generic solution near the Big-Bang).

Within that scheme the growth or decay of inhomogeneities according to their
equation of state can be easily inferred, at least qualitatively (see [10,
12]):
they decay when matter violates the strong energy condition, that is inflates,
and grow otherwise, that is when the effective adiabatic index $\Gamma_{eff}$
of
matter is $>2/3$. Therefore when matter is a scalar field the inhomogeneities
first grow ($\Gamma_{eff}=2$), then decay during the inflationary period, to
grow again at the end of inflation ($\Gamma_{eff}=1$), {\it at the
condition} that they do not grow so large during the first phase as to
prevent inflation to start.

The purpose of this paper is to give quantitative estimates on when inflation
may be halted by the presence of initial quasi-isotropic inhomogeneities. This
will be done by integrating numerically the ordinary second order
linear differential equations that govern their evolution in the long
wavelength approximation scheme.

\vskip 0.5 cm
\noindent
{\bf 2. The equations}
\vskip 0.5 cm

The long wavelength approximation scheme [10] consists in looking for solutions
of Einstein's equations for gravity coupled to a scalar field whose 3-metric
(in
the $t=Const$ slices of a synchronous reference frame) can be expanded as a sum
of spatial tensors of increasing order in the gradients of a ``seed" metric
with
 time dependent coefficients. The line element is thus of the form:
$$ds^2=-dt^2+\gamma_{ij}(t,x^k)dx^idx^j\quad ,\quad
\gamma_{ij}=^{(1)}\gamma_{ij}+  ^{(3)}\gamma_{ij}+ ^{(5)}\gamma_{ij}+...$$
$$^{(1)}\gamma_{ij}=a^2(t)h_{ij}(x^k)\quad ,\quad
^{(3)}\gamma_{ij}=a^2\left[ a_2(t)Rh_{ij}+b_2(t)R_{ij}\right]\eqno(1)$$
$$\eqalign{^{(5)}\gamma_{ij}=&a^2\left[a_4(t)R^2h_{ij}+b_4(t)RR_{ij}+
c_4(t)R_{lm}R^{lm}h_{ij}+d_4(t)R_{im}R_j^m\right]\cr
&+a^2\left[e_4(t)\nabla_m\nabla^mRh_{ij}+f_4(t)\nabla_i\nabla_jR+
g_4(t)\nabla_m\nabla^mR_{ij}\right],\cr}$$
and the scalar field of the form:
$$\varphi=\phi(t)+\phi_2(t)R+\phi_4(t)R^2+\psi_4(t)R_{lm}R^{lm}+
\xi_4(t)\nabla_m\nabla^mR+...\eqno(2)$$
This solution is built on a ``seed" $h_{ij}(x^k)$  of 6 functions of space
 which can be reduced to 3 by a suitable choice of spatial
coordinates. As shown in [10], it is not a generic solution (it should then
depend on 6 physically distinct functions) but it is an attractor of a class of
generic solutions. At lowest order the metric reduces to
$\gamma_{ij}=a^2(t)h_{ij}(x^k)$ and $\varphi=\phi(t)$, hence the name
``quasi-isotropic" given to the solution.
 $\nabla_i$ is the  covariant derivative with respect to
$h_{ij}$, $R_{ij}$  its  Ricci tensor, $R\equiv R^{ij}h_{ij}$ the curvature
scalar. The coefficients $a(t), a_2(t)$ etc are functions of time which are
determined by Einstein's equations. We shall  denote by $L$ the characteristic
comoving length on which the spatial metric varies:
$\partial_i\gamma_{jk}\simeq L^{-1}\gamma_{jk}$.

At zeroth order Einstein's equations reduce to the Friedmann equations for a
spatially flat Robertson-Walker (fRW) universe and determine $a(t)$ and
$\phi(t)$. Introducing the dimensionless variables:
$$T=mt\quad ,\quad F=2\sqrt{3\pi G}\phi\quad,\quad H={3\over m}{1\over a}{d
a\over dt} \quad ,\quad S=am^{1/3}\eqno(3)$$
where $m$ is the mass of the scalar field,
they read:
$${dS\over dT}={HS\over3}\quad ,\quad {dH\over dT}=-(H^2-F^2)\quad ,\quad
{dF\over dT}=-\sqrt{H^2-F^2}.\eqno(4)$$
The solutions of eqs (4) depend on three integration ``constants": the time
$\tilde T(x^k)$ of the Big-Bang that we shall restrict to be $\tilde T=0$ (see
[11] for an analysis of delayed Big Bang solutions); the size $\tilde S (x^k)$
of the scale factor at some given initial time: a ``constant" which can be
absorbed without loss of generality in a redefinition of the seed metric
$h_{ij}$; and finally  the initial value $\tilde F$ for the scalar field. A
priori $\tilde F$ depends on space but the first order Einstein equations (see
[10] for details): $\partial_iH=-4(dF/dT)\partial_iF$ imposes
that it does not. Its value, the same for all $x^k$, tells us which curve of
the
$(F, dF/dT)$ phase diagramme the solution follows, and hence determines the
total amount of inflation. We shall therefore integrate these equations
up to the beginning of inflation with, as initial conditions at
time $T=\epsilon$:
$$S_{in}=\epsilon^{1/3}\quad ,\quad
H_{in}={1\over\epsilon}\quad ,\quad F_{in}=\tilde F,\eqno(5)$$
$\epsilon$ being chosen such that $(\ln\epsilon)^2$ is numerically
negligible compared to $\epsilon^{-2}$. If matter were a perfect fluid with
adiabatic index $\Gamma$ the scale factor $S$ would grow as $T^{2/3\Gamma}$. In
the initial regime (5) the scalar field hence behaves like a stiff fluid
($\Gamma=2$) [1]. When $\tilde F$ is large enough the solution then enters an
inflationary phase characterized by a slow linear decrease of $F$ and a quasi
exponential growth of $S$. The comoving Hubble radius $L_H=3/SH$ hence first
increases as $T^{2/3}$ during the stiff fluid regime and then decreases
exponentially. The moment it reaches its maximum value can be taken as the
beginning of inflation. We shall denote $L^{inf}_H$ and $F_{inf}$ the values of
the Hubble radius and the scalar field at that moment.

At third order Einstein equations determine the time dependent coefficients
$a_2(t), b_2(t)$ and $\phi_2(t)$ in (1-2).  In terms of the variables defined
in
(3) and introducing:
$$B_2=b_2m^{4/3}\qquad ,\qquad A_2=a_2m^{4/3}\qquad ,\qquad
F_2=4\sqrt{{\pi G\over3}}m^{4/3}\phi_2\eqno(6)$$
they can be written  as (see [10]):
$${du\over dT}=-2S\qquad ,\qquad {dB_2\over dT}={u\over S^3}\eqno(7)$$
$${dv\over dT}=-{u\over 4S}{d\over dT}
\left[\left(S{dF\over dT}\right)^{-2}\right]\quad ,\quad
{dA_2\over dT}=v\left({dF\over dT}\right)^2\eqno(8)$$
$$F_2=-{1\over dF/dT}\left({1\over4}{dB_2\over dT}+{dA_2\over
dT}\right).\eqno(9)$$

It is easy to see that when the inflationary regime has set in, $B_2$ tends to
a constant (that we shall call $I_2$) as well as $F_2$. As for $A_2$ it
decreases
linearly in time. Now, as shown in [10], a gauge transformation modifies the
coefficient of the scalar curvature $R$ in ${}^{(3)}\gamma_{ij}$ but leaves
untouched the Ricci term. During inflation it amounts to adding an arbitrary
linear function of time to $A_2$. Therefore the linear decrease of $A_2$ can be
gauged away by a suitable choice of initial conditions for $A_2$, $v_2$,
leaving
$B_2$ as the only relevant quantity to be studied.
Physically the geometry evolves
from one configuration, $h_{ij}$, to another, $h_{ij}+(I_2/m^{4/3})R_{ij}$,
where
$I_2$ is the ``imprint" on the geometry left by the initial inhomogeneity. We
shall therefore integrate (7)
with the initial conditions (5) at $T=\epsilon$ together
with $dF/dT|_{in}=-1/\epsilon$ and:
$$u_{in}=-{3\over2}\epsilon^{4/3}\quad ,\quad B_2|_{in}
=-{9\over8}\epsilon^{4/3}. \eqno(10)$$

\vskip 0.5 cm
\noindent
{\bf 3. The solutions}
\vskip 0.5cm

Equations (4) with initial conditions (5) and equations (7) with initial
conditions (10) are integrated numerically. The results can be encapsulated in
a plot of $\tau_2\equiv\sqrt{|I_2|}/L^{inf}_H$ as a function of $F_{inf}$. See
the curve $k=0$ of Fig 1.
We see that as $F_{inf}$ increases, $\tau_2$ tends to a constant close to 1
(when
a brutal extrapolation of the analytical behaviours (5) and (10) would have
given
$1/2\sqrt 2$).
Going back to the expressions (1-2) for the metric and the scalar field, a
sufficient condition for inflation then appears clearly : inflation will
set in if the corrective terms $^{(3)}\gamma_{ij}$ remain small compared to
$^{(1)}\gamma_{ij}$ that is if $I_2/L^2<1$ which is equivalent, because
$\tau_2$ tends to 1, to $L>L_H^{inf}$. This condition   says
that the size $L$ of the inhomogeneity must be larger than the (comoving)
Hubble
radius at the onset of inflation, when the brutal analytic extrapolation gives
that it can be smaller by a factor $2\sqrt 2=2.8$.  Moreover, when the zeroth
order barely inflates, that is when $F_{inf}$ is small, $\tau_2=\alpha>1$, so
that a sufficient condition for inflation is $L>\alpha L_H^{inf}$.  We thus
recover in this semi-analytical approach the result of ref [8].

\vskip 0.5 cm
\noindent
{\bf 4. Convergence of the iteration scheme}
\vskip 0.5cm

A rigourous mathematical analysis of the convergence of the series (1-2) is
certainly beyond the scope of this paper. Using the results of the appendix of
ref [10] we can however compute the next order, that is the coefficients
$a_4, b_4$ etc and see if including them spoils the conclusion of the
preceeding
paragraph, drawn from the first iteration.

The relevant coefficients, which are not affected by a gauge transformation,
are $c_4$, $d_4$ and $g_4$. We shall concentrate on $g_4$ which satisfies the
following
linear second order differential equations:
$${dw\over dT}=B_2S\quad ,\quad {dG_4\over dT}=-{w\over S^3}\eqno (11)$$
where we have introduced $G_4\equiv m^{8/3} g_4$ with the initial conditions:
$$w_{in}=-{27\over64}\epsilon^{8/3}\quad ,\quad
G_4|_{in}=-{81\over8^3}\epsilon^{8/3}.\eqno (12)$$
(Knowing $G_4$ and $B_2$, the coefficient $D_4\equiv m^{8/3}d_4$, is readily
obtained : $D_4=-4G_4+{1\over2}B_2^2$; see [10].)

The results of the integration of (11-12) are summarized in the diagramme
$\tau_4\equiv |G_4|^{1/4}/L_H^{inf}$ as a function of $F_{inf}$ (see the curve
$k=0$ of Fig. 2). We see that $\tau_4$ follows the same pattern as $\tau_2$. As
$F_{inf}$ increases it tends a constant value of the order of $0.63$ when an
extrapolation of the analytical behaviour (16) would have given
$8^{-3/4}=0.21$. As for the ratio $|D_4|^{1/4}/L_H^{inf}$ it tends to $0.60$.

The inclusion of these fourth order terms therefore
does not spoil the conclusion of the previous paragraph and, since
$\tau_4<\tau_2$,  is an indication (although certainly not a proof!) that the
exact solution does not differ drastically from the second order  solution.

\vskip 0.5 cm
\noindent
{\bf 5. Curvature versus gradient effects
 and an improved scheme in the case of
spherical symmetry}
\vskip 0.5cm

In the case when the inhomegeneity is imposed to be spherically symmetric, the
seed line element $d\sigma^2=h_{ij}dx^idx^j$ can be written in the form:
$$d\sigma^2={dr^2\over1-kr^2}+r^2d\Omega\eqno(13)$$
where $d\Omega$ is line element on the sphere and where $k(r)$ is an arbitrary
function of $r$.

Expressing the Ricci tensor and its derivatives in function of $k(r)$ we can
write the long wavelength metric (1) and scalar field (2) under the form:
$$ds^2=-dt^2+R^2(C^2dr^2+r^2d\Omega)\eqno(14)$$
with
$$R=R_{cRW}^{app}\left[1+\left(a_2+{1\over4}b_2\right)k'r+...\right]
\quad ,\quad
R_{cRW}^{app}=a\left[1+\left(3a_2+b_2\right)k+...\right]\eqno(15)$$
$$C=C_{cRW}\left[1+\left(a_2+{1\over2}b_2\right)k'r+...\right]
\quad ,\quad C_{cRW}={1\over\sqrt{1-kr^2}}\eqno(16)$$
and
$$\varphi=\phi_{cRW}^{appr}+2k'r\phi_2+...\quad ,\quad
\phi_{cRW}^{app}=\phi+6k\phi_2+...\eqno(17)$$
where a prime denotes differentiation with respect to $r$ and where the time
dependent coefficients ($a,\phi$), ($a_2, b_2,\phi_2$) etc are the same as
before and satisfy eqns (3-5), (6-10), (11-12).

The rewriting of eq (1-2) under the form (14-17) shows clearly the two ways in
which a spherically symmetric inhomogeneity makes the solution depart from the
flat fRW solution. The first (trivial) effect is that of
curvature: if $k\neq0$ but all its derivative are taken to be zero, the long
wavelength solution (14-17) can be checked to be nothing but the Taylor
expansion in $t$ of the exact curved Robertson-Walker (cRW) solution. In this
case then the exact metric and scalar field are given by (14) with
$C=(1-kr^2)^{-1/2}$, and $R=a$ and $\phi$ satisfying the Friedmann equation:
$${dS\over dT}={HS\over3}\quad ,\quad {dH\over dT}=-(H^2-F^2)-{6K\over
S^2}\quad
,\quad {dF\over dT}=-\sqrt{H^2-F^2+9K/S^2}.\eqno(18)$$
where $T, F, H, S$ are defined by (3) and where $K=km^{-2/3}$.

Integrating (18) numerically with the same initial conditions as before
(eq(5)), we recover the results of ref [4], that is that if the curvature term
$K/S^2$ is too large then inflation is halted. This can be seen in Fig. 3 where
$L_H$ is plotted as a function of $T$ for different values of $K$ and a given
initial value $\tilde F$ for the scalar field. We see that a negative $K$
``favours" inflation whereas a positive value delays its setting in and, if
large enough, can even prevent it (dotted line).

The second effect is that of the gradients, that is the derivatives of $k(r)$,
which reflects the point to point correlation due to the variation of the
curvature. It can be enhanced by choosing a seed $k(r)$ such that it is
small everywhere but has a steep gradient $k'r$ around, say, $r=R$. From
(16) and the result of  section 3 that $\tau_2\simeq1$ we know that
these gradient effects will not prevent inflation to set in if the function
$k(r)$ is everywhere such that $k'r<2m^{4/3}/(L_H^{inf})^2$.

One can also improve the long wavelength scheme by
replacing in (14-17) the approximate cRW values by their exact values as given
by (18) and taking for $A_2,B_2,F_2$ the solutions of eq (7-9) where $S,H$ and
$F$ are taken to satisfy (18) instead of (4). The results are summarized in
Fig. 1 and 2 and show that the improved and the standard schemes coalesce for
large $F_{inf}$ that is for strongly inflating solutions. This confirms what
the previous section already indicated, that is that the iteration scheme seems
to converge nicely.

\vskip 0.5 cm
\noindent
{\bf 6. Conclusions}
\vskip 0.5cm

An important question in inflationary cosmology is: how generic is it? As
already shown by Goldwirth and Piran and confirmed here, inflation by itself
requires a certain level of homogeneity: it can start if initial
inhomogeneities
are larger than the local Hubble radius. While the numerical calculations of
[8]
can explore in a detailed way a specific (spherically symmetric) case of strong
initial inhomogeneity, the semi-analytical approach presented here is limited
to
rather small perturbations  but it gives a better global picture on the factors
that control the behaviour of the system. It gives only sufficient conditions
for the onset of inflation, not as strong as the necessary conditions obtained
in [8] but they are general and do not assume any spatial symmetry. We plan to
extend the comparison between these analytical and numerical approaches by
solving an identical initial value problem. This requires further investigation
of the various coordinates systems used.

\beginsection
Acknowledgments

 We thank Tsvi Piran and David Langlois for numerous discussions.
This work was partly supported by an ``Arc en Ciel, Keshnet" programme.

\vskip 1cm

\beginsection  References

[1] V.A. Belinski, I.M. Khalatnikov, JETP 36 (1973) 591

[2] see e.g. E.W.. Kolb, M.S. Turner, ``The Early Universe",
Addison-Wesley, 1990, A. Linde ``Particle Physics and Inflationary
Cosmology", Harwood, 1990

[3] V.A. Belinski, I.M. Khalatnikov, JETP 66 (1987) 441

[4] V.A. Belinski, H. Ishihara,  I.M. Khalatnikov, H. Sato, Progress Theor.
Phys. 79 (1988) 676

[5] D.S. Goldwirth, T. Piran, Physics Report 214 (1992) 223

[6] J. Kung and R. Brandenberger, Phys. rev. D 42 (1990) 1008; E. Calzetta and
M. Sakellariadou Phys. Rev D 45 (1992) 2802

[7] J.A. Shinkai, K. Maeda, Phys. Rev D, in press

[8] D.S. Goldwirth, T. Piran, Phys. Rev D 40 ( 1989) 3263; D.S. Goldwirth,
Phys. Rev. D 43 (1991) 3204

[9]  V.A. Belinski, E.M. Lifchitz, I.M. Khalatnikov, JETP 35 (1972) 838

[10] G.L. Comer, N. Deruelle, D. Langlois, J. Parry, Phys. Rev. D 49 (1994)
2759

[11] N.Deruelle , D.Langlois, to be submitted

[12] K. Tomita, N. Deruelle, accepted for publication by Phys. Rev. D
\vfill\eject
\beginsection
Figure Captions

{\bf Figure 1}
Plot of $\tau_2$ as a function of $F_{inf}$. The full line corresponds to
$k=0$, the dashed line above to negative $k$, the dashed dotted lines below to
positive $k$.
\vskip 0.5cm

{\bf Figure 2}
Plot of $\tau_4$ as a function of $F_{inf}$. The conventions are the same as in
Fig. 1.
\vskip 0.5cm

{\bf Figure 3}
Plot of the comoving Hubble radius as a function of time for various values of
the curvature term $k$ and a given initial value for the scalar field. The full
line corresponds to $k=0$, the dashed lines below to negative $k$, the
dashed dotted ones to positive $k$. Finally the dotted line is an example of a
curvature strong enough to prevent inflation from setting in.

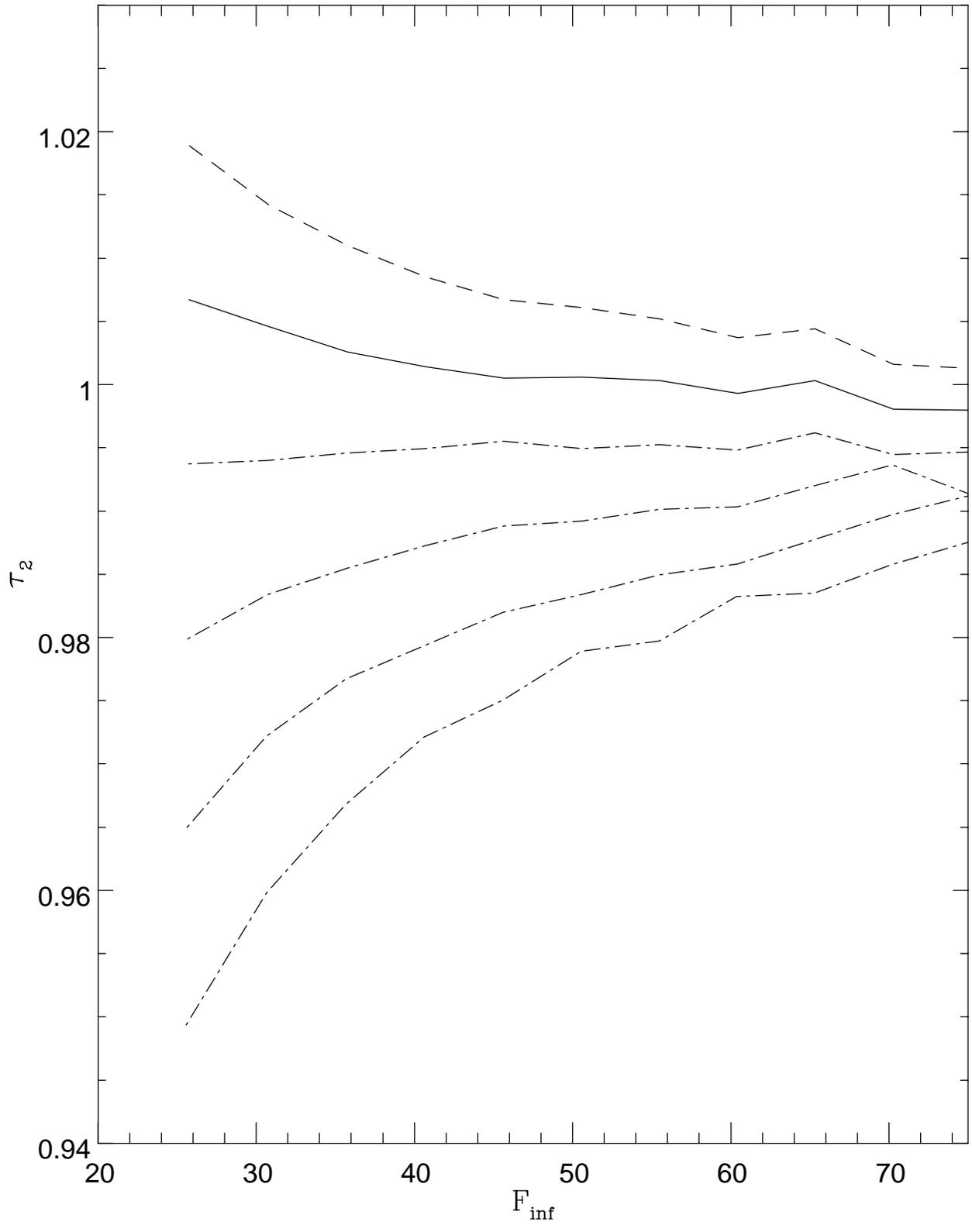
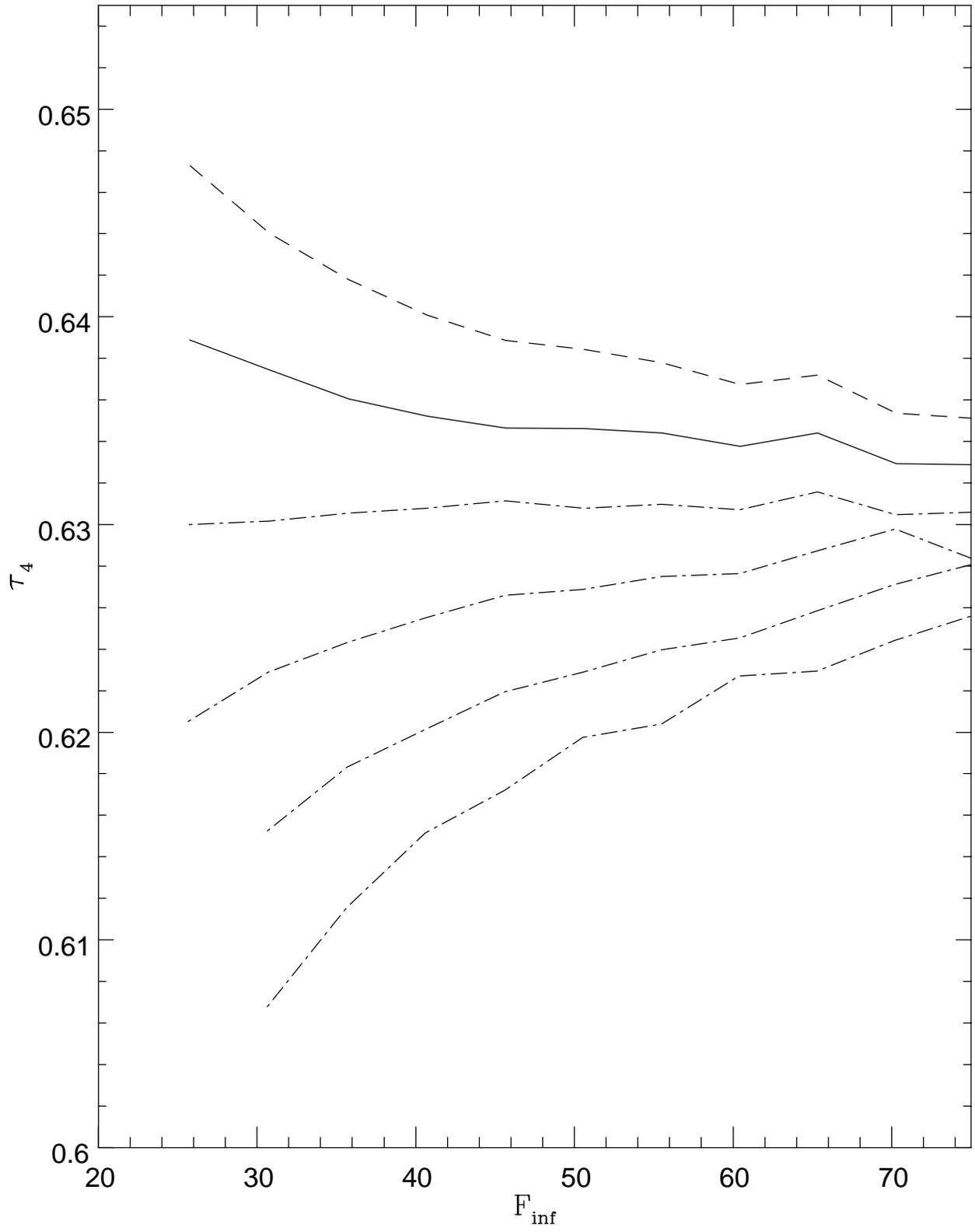
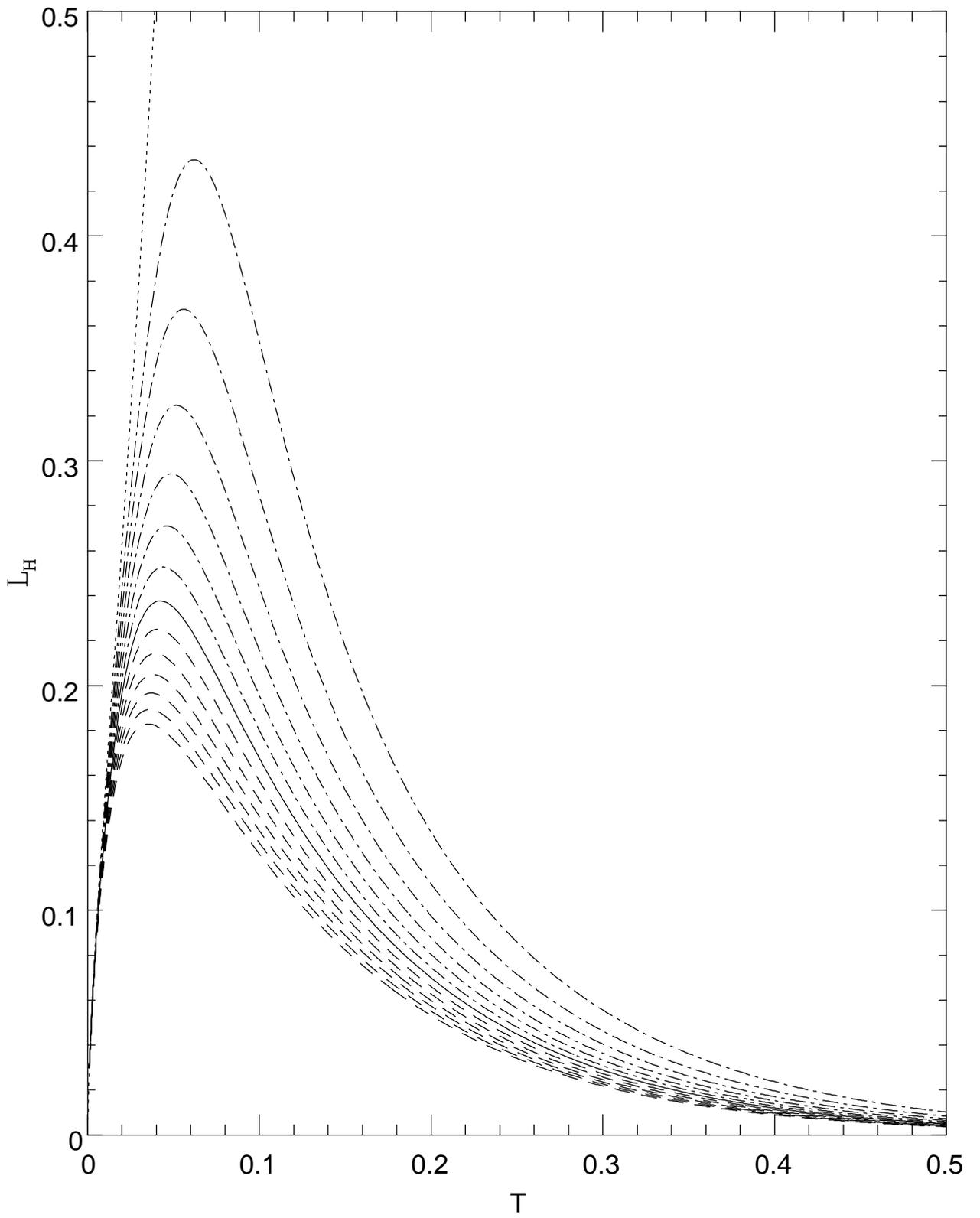
\end